\documentclass[lettersize,journal]{IEEEtran}
\usepackage{orcidlink}
\usepackage{amsmath,amsfonts}
\usepackage{algorithmic}
\usepackage{algorithm}
\usepackage{array}
\usepackage[caption=false,font=normalsize,labelfont=sf,textfont=sf]{subfig}
\usepackage{textcomp}
\usepackage{stfloats}
\usepackage{url}
\usepackage{verbatim}
\usepackage{graphicx}
\usepackage{cite}
\usepackage{multirow}
\begin{document}

\title{RV-IM100: Quantifying ISA Extension, Datapath Width, and Pipeline Depth Trade-offs in RISC-V Microarchitectures}

\author{Hyunwoo Kang\,\orcidlink{0009-0000-8804-5550},~\IEEEmembership{Student Member,~IEEE}
\thanks{Received DD Month 2026; revised DD Month 2026; accepted DD Month 2026. Date of publication DD Month 2026; date of current version DD Month 2026. (Corresponding author: Hyunwoo Kang.)}
\thanks{Hyunwoo Kang is with the Department of System Semiconductor Engineering, Sangmyung University, Cheonan, South Korea. (e-mail: hwctech1026@gmail.com)}
\thanks{Digital Object Identifier 00.0000/NN.2026.0000000}}

\markboth{IEEE Transactions on Very Large Scale Integration (VLSI) Systems}%
{Kang: RV-IM100}

\IEEEpubid{0000-0000~\copyright~ 2026 IEEE. All rights reserved, including rights for text and data mining, and training of artificial intelligence and similar technologies.}


\maketitle

\begin{abstract}
While functional RISC-V implementations are readily available in academia, controlled empirical studies that extend a single baseline architecture along multiple design axes and quantify the resulting trade-offs at each step remain scarce. This paper presents RV-IM100, a family of 10 incremental FPGA-implemented microarchitectures derived from a common 5-stage pipeline baseline, systematically varying datapath width from RV32 to RV64, instruction set from I to IM, and pipeline depth from 5 to 8~stages under controlled conditions. The I-to-IM extension produced strongly benchmark-dependent effects at the 5-stage level: CoreMark throughput more than doubled while Dhrystone throughput decreased marginally despite improved per-MHz efficiency. Within the RV32IM configuration, an iterative timing-closure methodology combined with pipeline deepening from 5 to 8 stages raised max frequency from 43 to 126\,MHz, increasing both Dhrystone and CoreMark throughput by 71\%, while per-MHz efficiency decreased by 41\%. The 6-to-7-stage transition caused throughput regression in RV64 despite higher frequency, revealing that the outcome depends on available frequency headroom. Cross-width comparison showed RV32 outperforming RV64 in absolute throughput, with per-MHz efficiency diverging by benchmark: RV64 led by 2.3\% in DMIPS/MHz while RV32 led by 4.6\% in CoreMark/MHz. At 8 stages, RV32 required 59\% fewer LUTs, 51\% fewer FFs, and 80\% fewer DSPs, indicating that the resource cost of width extension substantially exceeds the modest efficiency differences. These results provide a quantitative reference for design-space exploration in RISC-V microarchitectures. All RTL sources and benchmark configurations are publicly available.
\end{abstract}


\section{Introduction}
\IEEEPARstart{T}{he} RISC-V instruction set architecture (ISA) has experienced rapid adoption across both academia and industry since its introduction at UC~Berkeley~\cite{Waterman2011}. As a free and open-source ISA, RISC-V has enabled implementations ranging from educational pipelines~\cite{Patterson2020, Sodor, AhmadiPour2021} to configurable production cores~\cite{VexRiscv, Rocket2016, CVA62019} and high-performance out-of-order processors~\cite{Celio2015, Chen2020}. Comparative surveys~\cite{Dorflinger2021} have evaluated these implementations on common platforms, providing cross-core benchmarking data.
 
Despite this growing ecosystem, an important gap remains: the literature provides limited guidance on how to evolve a base design through successive architectural extensions while quantitatively documenting the trade-offs at each step~\cite{learn}. RISC-V's modular ISA specification supports extensive customization through optional extensions~\cite{Waterman2019}, yet few works provide stepwise methodologies with quantitative documentation of consequences at each stage. This gap manifests along three dimensions.
 
First, regarding \textit{bit-width extension}: while both RV32 and RV64 implementations exist independently~\cite{VexRiscv, picorv32, Rocket2016}, controlled comparisons measuring the resource cost, frequency impact, and performance implications of widening the datapath on otherwise identical microarchitectures have received little attention~\cite{Yongwoo2025}. Second, regarding \textit{instruction set extension}: existing implementations either include the M~extension as a fixed feature~\cite{Rocket2016, CVA62019} or support it as an optional configuration~\cite{Sodor, AhmadiPour2021, VexRiscv}, but the quantitative impact of adding multiplier and divider hardware across different pipeline depths remains underexplored. Third, regarding \textit{structural extension}: most implementations adopt a fixed pipeline depth, and empirical evidence for varying depth on the same base design remains limited. Recent work~\cite{Schoeberl2025} examined 3-to-5-stage organizations; our work extends this to 5-to-8~stages with ISA extension and cross-width analysis.
 
This paper presents RV-IM100, a family of FPGA-implemented RISC-V processor cores that systematically explores these three dimensions. Starting from a 5-stage MIPS-style pipeline implementing RV32I, we developed 10 incremental microarchitectures by progressively applying bit-width extension (RV32~to~RV64), instruction set extension (I~to~IM), and pipeline deepening (5~to~8~stages). Each variant was synthesized on a Xilinx Artix-7 FPGA under controlled conditions, with resource utilization measured via a methodology that externalizes memory interfaces to prevent synthesis tool constant propagation artifacts. This work makes the following contributions:
\begin{itemize}
\item An extensible ISA-driven design methodology for evolving a 5-stage RISC-V pipeline through bit-width, instruction set, and structural extensions up to 8~stages.
\item A timing closure methodology for FPGA frequency scaling, achieving 192\% frequency improvement (43~to~126\,MHz) for RV32IM and reaching 100\,MHz for RV64IM.
\item Quantitative characterization of all 10~variants across frequency, throughput, efficiency, resource utilization, and power, providing empirical data on cross-dimensional interactions.
\item Open-source release of all RTL sources and benchmark configurations.
\newpage
\end{itemize}
 
\section{Related Work}
 
Comparative evaluations of open-source RISC-V processors are well established. D\"orflinger et~al.~\cite{Dorflinger2021} compared Rocket~\cite{Rocket2016}, BOOM~\cite{Celio2015}, CVA6~\cite{CVA62019}, and SHAKTI~\cite{SHAKTI2016} under common implementation conditions in terms of performance, power, area, and efficiency. Their results highlighted both the substantial diversity among open-source application-class RISC-V processors and the importance of fair comparison under unified conditions. However, since the compared processors vary in multiple architectural features simultaneously, such studies are not well suited for isolating the effect of an individual design parameter. FPGA-oriented studies~\cite{Heinz2019, Holler2019} similarly provide valuable benchmarking references for ecosystem-level comparison, but because the evaluated processors are independent designs, they are less effective for isolating the structural consequences of a specific extension within a single incrementally evolved architecture.
 
ISA extension has been examined through the RVCoreP family: Islam et~al.~\cite{Islam2020} added the M~extension and Kanamori et~al.~\cite{Kanamori2021} added the C~extension to the same RV32I baseline~\cite{Hiromu2020}. However, these studies report on a single pipeline depth and width, without examining how M-extension cost scales with pipeline depth or datapath width.
 
Structural extension of a common baseline remains limited. Schoeberl's Wildcat~\cite{Schoeberl2025} compared pipeline organizations and reported that deeper pipelines do not necessarily yield higher frequency~\cite{Hartstein2002}, a finding highly relevant to the present work. Related structural optimization studies~\cite{Matthews2022, Zheng2022} address arithmetic-unit and area-performance trade-offs but do not systematically vary pipeline depth. Methodological frameworks such as BRISC-V~\cite{BRISC-V2019} and SCOoOTER~\cite{SCOoOTER2025} support exploration but generally provide frameworks rather than stepwise analysis of one processor lineage.
 
In contrast, this paper studies ten microarchitectures derived from a common baseline and extended incrementally along three axes, enabling more controlled comparison of each change's effect than cross-design studies.
 
\section{Architecture Design}
\label{sec:architecture}
 
\subsection{Design Philosophy}
\label{sec:philosophy}
The base architecture is derived from the RV32I\_Zicsr 5-stage pipeline presented in our prior work~\cite{Hyunwoo2025}. The present work adopts three additional design rules: (1)~the baseline architecture is preserved unless modification is strictly necessary for correct execution, enabling controlled comparison; (2)~extensions prioritize frequency-oriented optimization, targeting 100\,MHz for RV64; and (3)~RV32 variants are derived by reducing functionality from the completed RV64 design, ensuring identical pipeline control logic, forwarding paths, and hazard detection, so that observed differences reflect datapath width alone.
 
\subsection{Base Pipeline}
The base microarchitecture, designated 46F5SP~\cite{Hyunwoo2025}, is a traditional 5-stage pipeline (IF, ID, EX, MEM, WB) with 18~functional modules, 4~pipeline registers, exception handling, 2-bit branch prediction~\cite{Smith1981}, and full data forwarding. The designation \textit{46F} indicates 46~supported instructions.
 
\begin{figure}[!t]
  \centering
  \includegraphics[width=\linewidth]{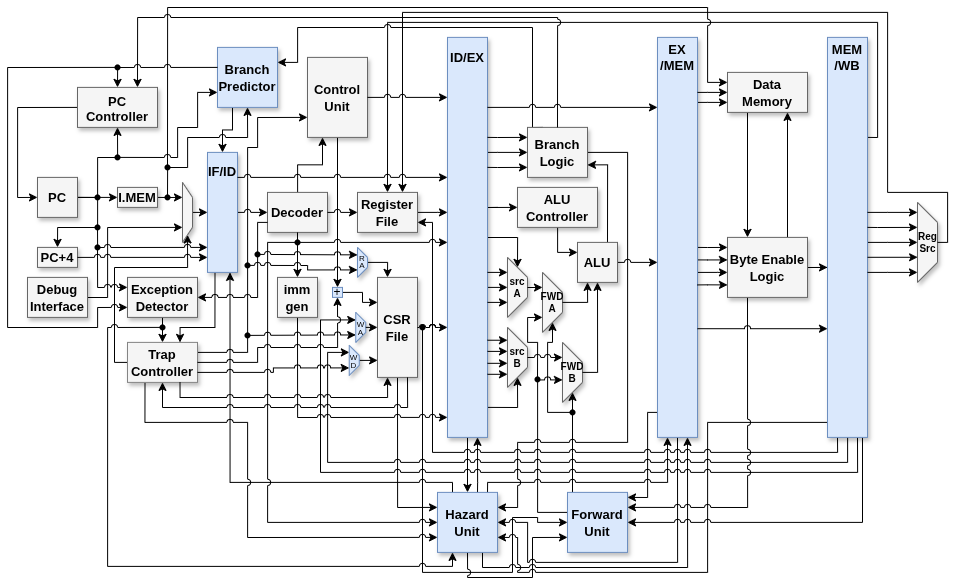}
  \caption{Simplified block diagram of the 46F5SP base microarchitecture.}
  \label{fig:5sp}
\end{figure}
 
\subsection{RV32-to-RV64 Parameterization}
\label{sec:rv32to64}
The RV64I architecture (59F5SP) is derived from 46F5SP by widening the datapath and register file via the \texttt{XLEN} parameter. Beyond the parametric width change, four modules require ISA-level modifications: the Instruction Decoder and Control Unit recognize two new opcodes (\texttt{OP-IMM-32}, \texttt{OP-32}) for W-suffix instructions; the ALU Controller accepts a 6-bit shift-amount field; the ALU is restructured into parallel 32-bit and 64-bit datapaths with result selection based on instruction width; and the Byte Enable Logic supports doubleword operations (\texttt{LD}, \texttt{SD}, \texttt{LWU}). No additional modules are introduced.
 
\subsection{I-to-IM Extension}
The ALU is extended with dedicated multiplier and divider units in the same dual-width structure, with the top-level ALU selecting the appropriate result based on the instruction width. The multiplier is implemented as a 3-stage pipelined unit that leverages FPGA DSP primitives: the first stage registers operands and resolves their signs to unsigned absolute values, the second stage performs the unsigned multiplication via DSP blocks inferred through the \texttt{use\_dsp} synthesis attribute, and the third stage applies sign correction. For the DWORD variant, a single 64$\times$64 multiplication is decomposed into four parallel 32$\times$32 DSP multiplications whose partial products are accumulated in the third stage. The divider implements a restoring division algorithm using a combined remainder-quotient shift register; a 4-state FSM controls the operation, with the CALCULATE state iterating for $N$~cycles (32~for WORD, 64~for DWORD), performing one trial-subtraction and shift per cycle. The ALU Controller is extended to generate \texttt{mul\_start} and \texttt{div\_start} signals upon decoding M-extension opcodes, with in-flight tracking registers preventing re-triggering while an operation is in progress. During multi-cycle operations, the Hazard Unit stalls the entire pipeline until completion.
 
\subsection{6-Stage Pipeline}
The 6-stage pipeline (72F6SP) is designed to break the critical path present in the 5-stage organization. In 46F5SP, the EX~stage performs ALU computation and branch resolution within a single clock cycle: the forwarded operands enter the ALU, the ALU produces a result and a zero flag, and the Branch Logic immediately consumes these outputs to determine whether a branch is taken and to compute the corrected program counter. This serial chain from forwarding mux through ALU to PC~update constitutes the longest combinational path in the 5-stage design. To shorten this path, the ALU output is registered at a new pipeline boundary (EX\_BR\_Register), splitting the original EX~stage into EX~(ALU computation) and BR~(branch evaluation using the registered ALU result and zero flag). The pipeline becomes IF, ID, EX, BR, MEM, WB. Because branch resolution is deferred by one stage, the misprediction penalty increases from two to three flush cycles, reducing IPC on branch-intensive code.
 
\subsection{7-Stage Pipeline}
The 7-stage pipeline (72F7SP) inserts an Instruction Out stage~(IO) between IF and ID, yielding IF, IO, ID, EX, BR, MEM, WB. Unlike the 5-to-6-stage transition, which restructured the execution back-end, this change addresses the instruction fetch front-end. In the 6-stage design, instruction memory uses distributed RAM with combinational read; to improve timing, it is migrated to synchronous Block~RAM (BRAM), which provides favorable timing characteristics but inherently requires one clock cycle of read latency. The IO~stage absorbs this latency; without it, every fetch would incur a structural hazard stall, negating the frequency benefit. The data memory is likewise migrated to synchronous BRAM, but requires no additional stage: the registered ALU result in the BR~stage serves as an early address presentation one cycle before MEM, naturally absorbing the BRAM read latency while preserving the existing single-cycle load-use penalty. For store instructions, a one-cycle \texttt{write\_done} stall ensures synchronous BRAM write completion before any subsequent load to the same address.

This front-end deepening has a larger structural IPC cost than the earlier 5-to-6-stage transition. The misprediction penalty increases from three to four flush cycles, but the more significant cost is qualitative: the fetched instruction must now traverse both IF and IO before reaching ID, imposing a two-cycle front-end refill latency that doubles the recovery cost after every misprediction or taken jump. The resulting IPC impact depends on whether the frequency gain from the BRAM migration is sufficient to compensate; this trade-off is evaluated quantitatively in Section~\ref{sec:results}.
 
\subsection{8-Stage Pipeline}
\label{sec:8stage}
The 8-stage pipeline (72F8SP) inserts an Execution Ready stage~(EXR) between ID and EX, yielding IF, IO, ID, EXR, EX, BR, MEM, WB. Like the 5-to-6-stage transition, this change targets the execution back-end. In the 7-stage design, the EX~stage performs hazard detection, forwarding data selection, source multiplexing, and ALU computation within a single cycle, forming a four-stage serial combinational chain that constitutes the critical path. The EXR stage breaks this chain by moving hazard detection, forwarding selection, and source multiplexing into a separate stage where the resolved operands are registered. The EX~stage then receives pre-resolved operands and performs only the ALU computation, shortening the critical path from the full hazard-to-ALU chain to the longer of two independent shorter segments.
 
The EXR stage also creates a new data hazard absent from the 7-stage design. In the 7-stage pipeline, a producing instruction's ALU result is available in BR when the consuming instruction arrives at EX for forwarding. In the 8-stage pipeline, forwarding resolution has moved to EXR, so the producing instruction is still in EX and the consuming instruction enters EXR, the ALU result is not yet available as a forwarding source. The Hazard Unit must stall for one cycle, allowing the producing instruction to advance to BR before the consuming instruction resolves its operands in EXR. Unlike the load-use hazard, this \textit{execution-use hazard} applies to all back-to-back register dependencies, introducing a per-occurrence one-cycle penalty that did not exist in shallower configurations. The misprediction penalty increases from four to five flush cycles, while the front-end refill latency remains at two cycles. Whether the frequency improvement from breaking the forwarding-ALU critical path offsets these penalties is evaluated in Section~\ref{sec:results}.
 
\begin{figure*}[!t]
  \centering
  \includegraphics[width=0.85\textwidth, height=0.45\textheight, keepaspectratio]{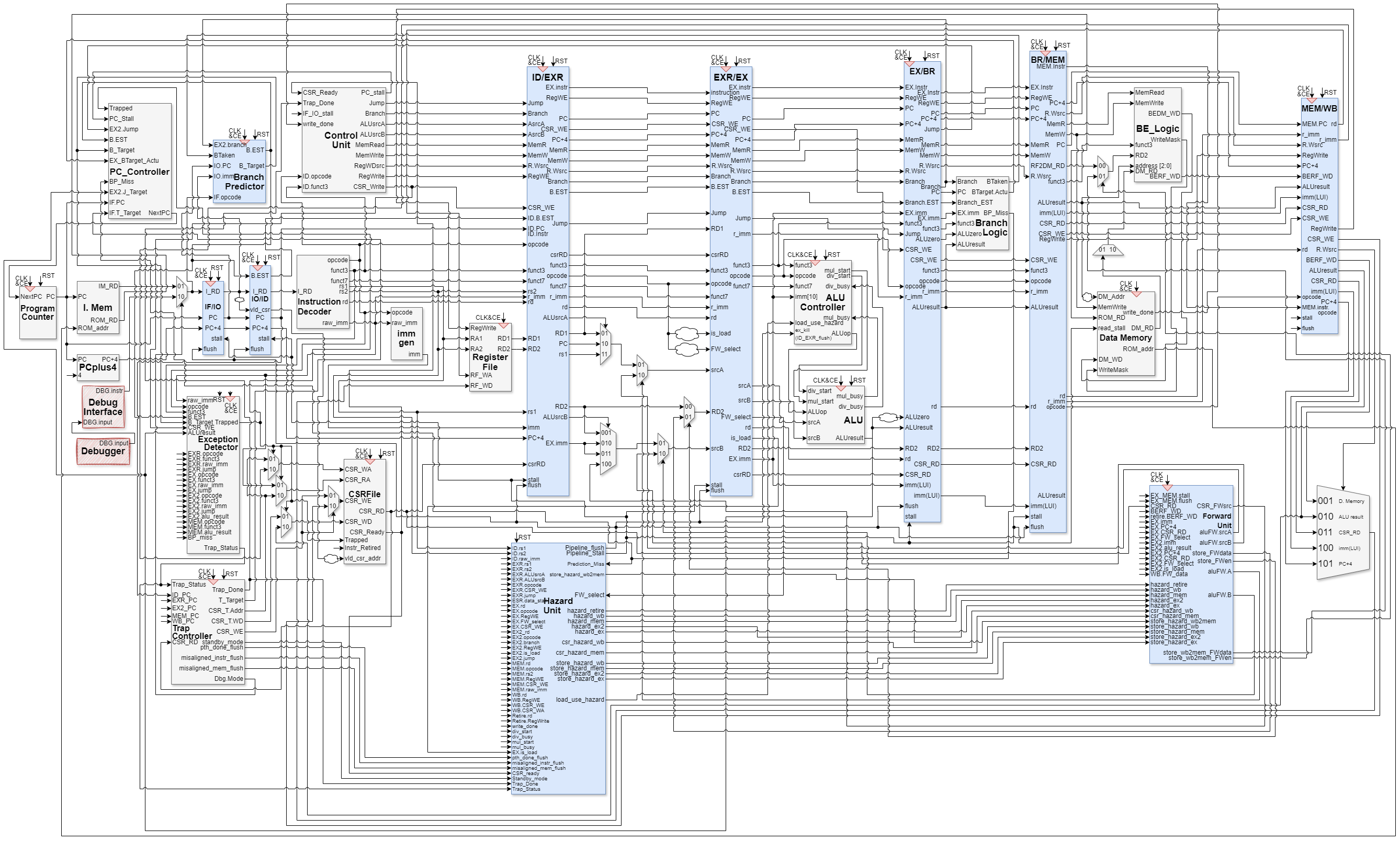}
  \caption{Signal-level block diagram of 72F8SP, the final RV64IM 8-stage pipeline microarchitecture (IF, IO, ID, EXR, EX, BR, MEM, WB).}
  \label{fig:rv64im8sp}
\end{figure*}
 
\section{Timing Closure Methodology}
\label{sec:timing_closure}
 
Pipeline deepening addresses timing at the structural level, but each configuration also requires RTL-level optimization to eliminate secondary critical paths. The timing closure process follows a three-phase iterative approach: (1)~analyze post-implementation timing reports to identify the dominant critical path; (2)~apply targeted RTL modifications without altering pipeline structure; (3)~evaluate whether remaining paths necessitate stage insertion.
 
Static timing analysis on the initial 7-stage SoC (operating at 55.9\,MHz, Table~\ref{tab:fmax_progression}) under a 10\,ns constraint identified the critical path as a single-cycle chain in the EX~stage with a total delay of 16.1\,ns (logic 5.0\,ns, net 11.1\,ns), with net delay accounting for 69\% of the total and indicating that high fan-out routing on forwarding control signals was the primary bottleneck. Twelve RTL optimizations were applied iteratively across three categories: \textit{forwarding network optimizations}, \textit{pre-computation} (relocating operations to earlier stages where inputs were already registered), and \textit{deferral} (moving operations to later stages as deliberate trade-offs). Table~\ref{tab:timing_summary} lists the individual optimizations and the cumulative effect; the overall trend is downward, although individual steps occasionally show net delay increases due to placement variation across synthesis runs. The three most effective optimizations were MEM forwarding data pre-registration ($-2.394$\,ns), which eliminated the MEM-stage opcode decode from the forwarding data selection path; WB forwarding data pre-registration ($-0.942$\,ns); and one-hot multiplexer conversion ($-0.739$\,ns), which flattened the cascaded 4-level forwarding multiplexer to a single wide-OR level, reducing the forwarding fan-out from 177 to 88.
 
\begin{table}[!t]
\centering
\caption{Cumulative timing closure results for the 7-stage pipeline.}
\label{tab:timing_summary}
\footnotesize
\begin{tabular}{llccc}
\hline
\textbf{Category} & \textbf{Optimization} & \textbf{Logic} & \textbf{Net} & \textbf{Total} \\
 & & \textbf{(ns)} & \textbf{(ns)} & \textbf{(ns)} \\
\hline
& Initial 7-stage & 5.0 & 11.1 & 16.1 \\
\hline
\multirow{7}{*}{Forwarding}
& One-hot MUX conv. & 4.1 & 11.2 & 15.4 \\
& Retire fwd. elim. & 4.9 & 10.1 & 15.0 \\
& CSR fwd. elim. & 5.7 & 9.0 & 14.7 \\
& BR fwd. pre-reg. & 4.0 & 10.4 & 15.4 \\
& MEM fwd. data pre-reg. & 3.4 & 9.6 & 13.0 \\
& WB fwd. data pre-reg. & 4.5 & 7.6 & 12.1 \\
& Encoded MUX restruct. & 3.2 & 9.2 & 12.4 \\
\hline
\multirow{3}{*}{Pre-computation}
& MMIO cmp. (MEM$\to$BR) & 5.4 & 6.6 & 12.0 \\
& CSR valid. (ID$\to$IO) & 3.1 & 8.6 & 11.6 \\
& Hazard sig. (EX$\to$ID) & 3.1 & 8.8 & 11.9 \\
\hline
\multirow{3}{*}{Deferral}
& Exc.Det. (EX$\to$BR) & 4.8 & 7.2 & 12.0 \\
& JALR res. (EX$\to$BR) & 3.4 & 8.4 & 11.9 \\
& ALU sign-ext. (EX$\to$BR) & 2.8 & 8.5 & 11.4 \\
\hline
& \textbf{Final 7-stage} & \textbf{2.8} & \textbf{8.5} & $\sim$\textbf{11.4} \\
\hline
\end{tabular}
\end{table}
 
The cumulative effect was a 29\% reduction in data path delay (16.1~to~11.4\,ns). The residual critical path consisted of three serially irreducible components: hazard detection ($\sim$2\,ns), forwarding and source selection ($\sim$2\,ns), and 64-bit ALU computation ($\sim$4\,ns). Including $\sim$2.5\,ns clock overhead, the minimum period was constrained to $\sim$10.5\,ns. Closing the gap to 10\,ns (100\,MHz) required inserting the EXR stage (Section~\ref{sec:8stage}). Table~\ref{tab:fmax_progression} summarizes the three-phase frequency progression from the baseline 7-stage SoC through RTL optimization to the final 8-stage design.

\begin{table}[!t]
\centering
\caption{Frequency progression through timing closure phases (RV64IM, Artix-7).}
\label{tab:fmax_progression}
\footnotesize
\begin{tabular}{lrl}
\hline
\textbf{Phase} & \textbf{$F_\mathrm{max}$} & \textbf{Description} \\
\hline
72F7SP initial SoC & 55.9\,MHz & before RTL optimization \\
72F7SP after RTL optimization & $\sim$88\,MHz & 12 optimizations applied \\
72F8SP after EXR insertion & 100.8\,MHz & RTL optimization + EXR \\
\hline
\end{tabular}
\end{table}

A common principle underlies both pre-computation and deferral: an operation need not be evaluated in the pipeline stage where its result is consumed. Advancing computations to stages where inputs are already available, or deferring them to stages with less timing pressure, enables workload balancing without modifying pipeline structure.
 
\section{Experimental Setup}
\label{sec:experimental_setup}
 
\begin{figure}[!t]
  \centering
  \includegraphics[width=\linewidth]{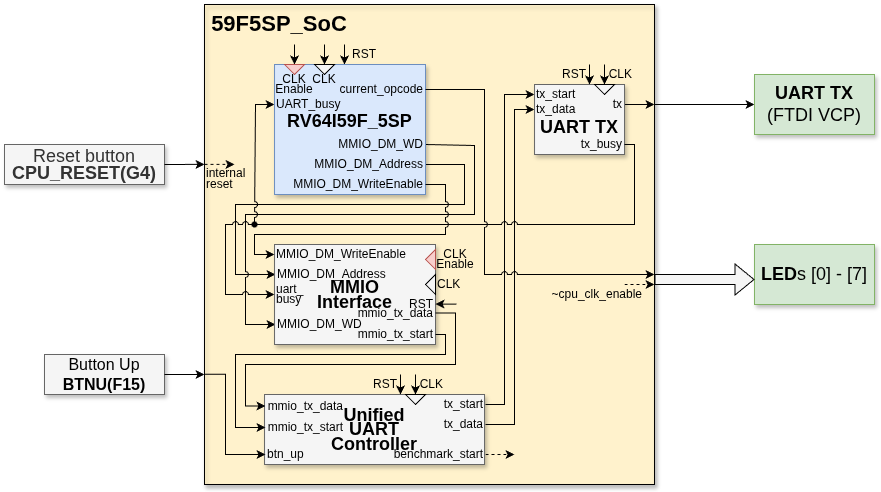}
  \caption{Signal-level block diagram of the RV-IM100 SoC.}
  \label{fig:soc}
\end{figure}

All designs are implemented on a Digilent Nexys Video board with a Xilinx Artix-7 XC7A200T-1SBG484C FPGA ($-1$ speed grade). Each processor core is integrated into an SoC wrapper, shown in Fig.~\ref{fig:soc}, providing BRAM-based instruction and data memory, UART transmitter, MMIO interface, PLL-based clock generator, and GPIO.
 
Synthesis and implementation use AMD Vivado 2025.2 with Performance\_Explore and Performance\_ExplorePostRoutePhysOpt strategies. RTL is written in Verilog HDL with functional verification via Icarus Verilog and GTKWave.
 
Two synthesis configurations are used. \textit{Core-only} measurement externalizes memory interfaces to prevent constant propagation artifacts, with a 5\,ns timing constraint to report $F_\mathrm{max}$. \textit{SoC} measurement includes complete memory and reports the achievable operating frequency.
 
Benchmarks are compiled using \texttt{riscv64-unknown-elf-gcc}~15.2.0. Dhrystone~2.1~\cite{Weicker1984} uses \texttt{-O2} with 300\,000 iterations; CoreMark~\cite{CoreMark} uses \texttt{-O2 -fno-common -funroll-loops}.
 
\begin{table}[!t]
\centering
\caption{The ten RV-IM100 microarchitecture variants.}
\label{tab:variants}
\footnotesize
\begin{tabular}{llcc}
\hline
\textbf{Designation} & \textbf{ISA} & \textbf{XLEN} & \textbf{Stages}\\
\hline
46F5SP & RV32I\_Zicsr & 32 & 5 \\
54F5SP & RV32IM\_Zicsr & 32 & 5 \\
54F6SP & RV32IM\_Zicsr & 32 & 6 \\
54F7SP & RV32IM\_Zicsr & 32 & 7 \\
54F8SP & RV32IM\_Zicsr & 32 & 8 \\
59F5SP & RV64I\_Zicsr & 64 & 5 \\
72F5SP & RV64IM\_Zicsr & 64 & 5 \\
72F6SP & RV64IM\_Zicsr & 64 & 6 \\
72F7SP & RV64IM\_Zicsr & 64 & 7 \\
72F8SP & RV64IM\_Zicsr & 64 & 8 \\
\hline
\end{tabular}
\end{table}
 
\section{Results}
\label{sec:results}
 
\subsection{Frequency Scaling}
 
\begin{figure}[!t]
  \centering
  \includegraphics[width=\linewidth]{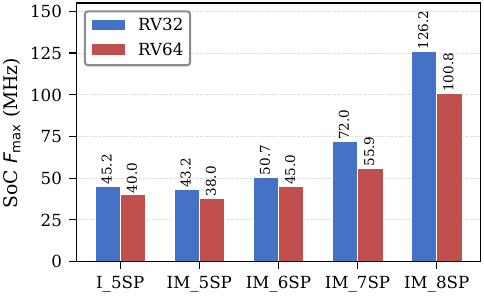}
  \caption{SoC maximum operating frequency across pipeline variants.}
  \label{fig:soc_fmax}
\end{figure}

Fig.~\ref{fig:soc_fmax} shows the SoC $F_\mathrm{max}$ for all variants. The base I\_5SP operates at 45.2\,MHz (RV32) and 40.0\,MHz (RV64). The M~extension marginally reduces frequency to 43.2 and 38.0\,MHz, representing the lowest operating points.
 
From IM\_5SP to IM\_6SP, the EX/BR split raises frequency to 50.7\,MHz (RV32) and 45.0\,MHz (RV64). The 6-to-7 transition reaches 72.0 and 55.9\,MHz via the combined insertion of the IO stage and migration of instruction memory from distributed RAM to synchronous BRAM, eliminating the combinational read path from the timing budget. The most substantial improvement appears at IM\_8SP: 126.2\,MHz (RV32) and 100.8\,MHz (RV64), reflecting the combined effect of the twelve RTL-level timing closure optimizations described in Section~\ref{sec:timing_closure} and the EXR stage insertion that separates the forwarding network from the ALU computation. Overall, from the lowest point at IM\_5SP to the final IM\_8SP configuration, RV32 achieves a 192\% frequency improvement (43.2$\to$126.2\,MHz) and RV64 achieves a 165\% improvement (38.0$\to$100.8\,MHz), with the pipeline organization evolving from IF, ID, EX, MEM, WB to IF, IO, ID, EXR, EX, BR, MEM, WB.
 
RV32 consistently achieves higher $F_\mathrm{max}$, with the gap widening from $\sim$5\,MHz through IM\_6SP to 25\,MHz at IM\_8SP, consistent with the narrower datapath producing shorter carry chains and lower routing congestion.
 
\subsection{Benchmark Performance}
 
\subsubsection{Absolute Throughput}
The two benchmarks respond differently to the I-to-IM transition. In Dhrystone, M~extension hardware does not improve throughput; RV32 decreases from 87\,378 to 83\,820~Dhrystones/s and RV64 from 80\,482 to 76\,767, primarily due to the accompanying 2\,MHz frequency reduction at the 5-stage level. This indicates that integer multiplication and division operations account for a negligible fraction of Dhrystone execution time. In contrast, CoreMark throughput increases substantially (RV32: 50$\to$117~iterations/s; RV64: 37$\to$91), reflecting its heavier reliance on multiplication-intensive matrix and list operations.
 
The 6-to-7-stage transition produces a notable exception in RV64: Dhrystone throughput drops from 87\,548 to 78\,124 and CoreMark stagnates at 100~iterations/s despite a frequency increase from 45.0 to 55.9\,MHz. This regression suggests that, in the RV64 7-stage configuration, the two-cycle refill latency and wider flush window erode throughput more than the frequency gain compensates. RV32 does not exhibit this regression, as its higher frequency (72.0\,MHz) provides sufficient headroom. This observation suggests that the viability of a given pipeline deepening strategy depends not only on the penalty it introduces but also on the absolute frequency headroom available.
 
At IM\_8SP, RV32 reaches 143\,668~Dhrystones/s (+64.4\% over I\_5SP) and 200~CoreMark iterations/s (+300\%), while RV64 reaches 117\,508~Dhrystones/s and 153~CoreMark iterations/s. The 7-to-8-stage transition targets the execution back-end and achieves the largest frequency improvement in the entire design space (54.2\,MHz for RV32, 44.9\,MHz for RV64), more than compensating for the execution-use hazard and the incremental misprediction penalty. The contrasting outcomes of the 6-to-7 and 7-to-8 transitions illustrate that pipeline deepening does not produce uniform throughput effects across configurations. The asymmetry is consistent with a structural distinction between front-end stages that add both flush penalty and refill latency and back-end stages that add only flush penalty, though further investigation would be needed to quantify each factor independently.
 
\begin{figure*}[!t]
\centering
\subfloat[Dhrystones/s]{\includegraphics[width=0.48\textwidth]{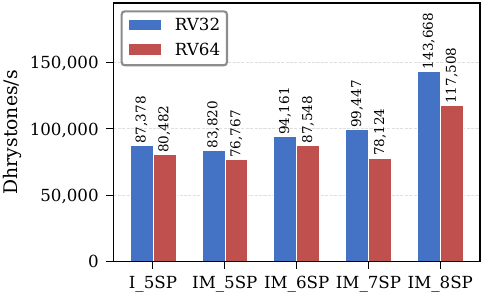}%
\label{fig:dhry_abs}}
\hfil
\subfloat[CoreMark iterations/s]{\includegraphics[width=0.48\textwidth]{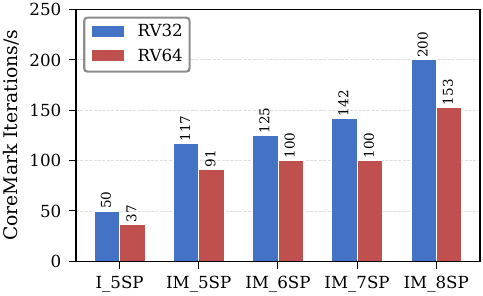}%
\label{fig:cm_abs}}
\caption{Benchmark absolute throughput across pipeline variants.}
\label{fig:benchmark_abs}
\end{figure*}
 
\subsubsection{Relative Efficiency}
Per-MHz efficiency, which isolates microarchitectural IPC from frequency effects, peaks at IM\_5SP and declines monotonically with each subsequent pipeline deepening. This is expected, as each additional stage increases the branch misprediction penalty and, in the case of IM\_8SP, the execution-use hazard introduced by the EXR stage adds a further one-cycle stall to consecutive register dependencies. DMIPS/MHz declines 41.0\% for RV32 (1.109$\to$0.654) and 41.8\% for RV64 (1.150$\to$0.669). CoreMark/MHz declines 41.2\% for RV32 (2.721$\to$1.600) and 36.1\% for RV64 (2.395$\to$1.530). Despite this efficiency loss, the absolute throughput gains at IM\_8SP confirm that frequency scaling more than compensates for the IPC degradation in all cases except the RV64 6-to-7 transition.

Cross-width comparison reveals asymmetric efficiency divergence: RV64 leads by 2.3\% in DMIPS/MHz at IM\_8SP (0.669~vs.~0.654), while RV32 leads by 4.6\% in CoreMark/MHz (1.600~vs.~1.530), with the gap widening to 8.5\% at IM\_7SP. The per-MHz gap suggests an IPC difference that cannot be attributed to cycle time alone; possible contributing factors include the additional W-suffix instructions generated by the RV64 compiler for 32-bit-typed CoreMark kernels and the wider immediate/address computations that increase dynamic instruction count, though confirming this hypothesis would require instruction-level profiling, which is left to future work.
 
\begin{figure*}[!t]
\centering
\subfloat[DMIPS/MHz]{\includegraphics[width=0.48\textwidth]{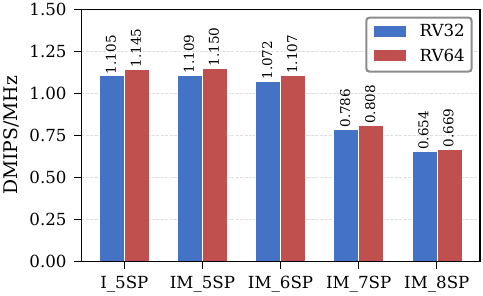}%
\label{fig:dmips_mhz}}
\hfil
\subfloat[CoreMark/MHz]{\includegraphics[width=0.48\textwidth]{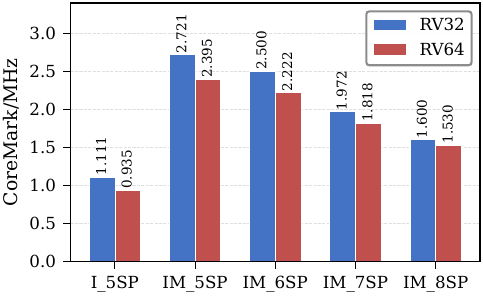}%
\label{fig:coremark_mhz}}
\caption{Benchmark relative efficiency across pipeline variants.}
\label{fig:benchmark}
\end{figure*}
 
\subsection{Resource Utilization}
 
\begin{figure*}[!t]
\centering
\subfloat[LUT]{\includegraphics[width=0.48\textwidth]{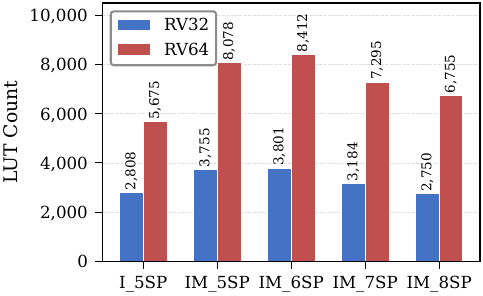}%
\label{fig:lut}}
\hfil
\subfloat[FF]{\includegraphics[width=0.48\textwidth]{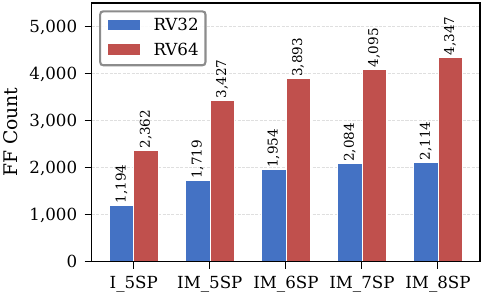}%
\label{fig:ff}}
\caption{Core-only resource utilization across pipeline variants.}
\label{fig:resource}
\end{figure*}

Fig.~\ref{fig:resource} shows core-only LUT and FF counts. DSP usage is 0~for I-extension variants and fixed at 4~(RV32) and 20~(RV64) for all IM~variants.
 
LUT utilization follows a non-monotonic pattern: the I-to-IM transition produces the largest single increase (RV32: 2\,808$\to$3\,755; RV64: 5\,675$\to$8\,078), corresponding to the dual-width ALU, multiplier, and divider hardware. The 5-to-6-stage transition adds a modest number of LUTs for the additional pipeline register, but from IM\_6SP onward LUT counts decrease at each subsequent stage, because pipeline stage insertion converts combinational logic paths into registered sequential logic, shifting resource consumption from LUTs to flip-flops. At IM\_8SP, the RTL-level timing closure optimizations contribute a further reduction by pre-registering forwarding values and eliminating redundant combinational paths, so both RV32 (2\,750) and RV64 (6\,755) are below their IM\_5SP baselines. FF~counts increase monotonically with pipeline depth, at approximately 200~FFs per additional stage for RV32 and 300--450 for RV64, proportional to the wider datapath signals that must be registered at each pipeline boundary.
 
At IM\_8SP, RV32 uses 59.3\% fewer LUTs (2\,750~vs.~6\,755) and 51.4\% fewer FFs (2\,114~vs.~4\,347) than RV64. The resource scaling is super-linear with respect to the 2$\times$ width increase because the overhead extends beyond the datapath itself to the forwarding network, hazard detection comparators, pipeline registers, and byte-enable logic, all of which must accommodate the wider data and address paths. For workloads not requiring 64-bit address space or data width, RV32 therefore offers a substantially more favorable cost-performance trade-off on resource-constrained FPGA targets.
 
\subsection{Power Estimation}
 
\begin{figure}[!t]
  \centering
  \includegraphics[width=\linewidth]{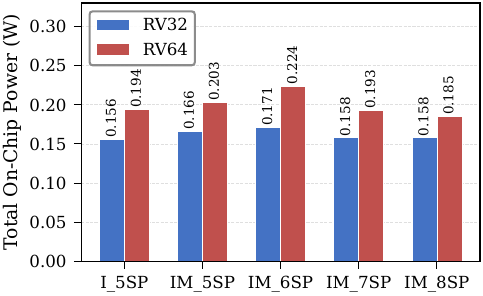}
  \caption{Core-only estimated total on-chip power across pipeline variants.}
  \label{fig:power}
\end{figure}

Fig.~\ref{fig:power} shows core-only estimated total on-chip power. RV64 consistently consumes more power than RV32, with a 17.1\% gap at IM\_8SP. Both width variants show a similar non-monotonic trend: power increases through IM\_6SP with M~extension hardware and the EX/BR stage split, then decreases at IM\_7SP and IM\_8SP. The subsequent decrease is consistent with the LUT-to-FF resource shift observed in Fig.~\ref{fig:resource}: as pipeline insertion converts combinational paths to registered logic, the proportion of flip-flops relative to LUTs increases, and flip-flops generally exhibit lower dynamic power than equivalent combinational LUT chains due to reduced glitching and more predictable switching activity. At IM\_8SP, RV64 consumes 17.1\% more power than RV32. These figures are Vivado post-implementation estimates based on default switching activity assumptions rather than simulation-derived activity data, and should be interpreted as relative comparisons between variants rather than absolute power measurements.
 
\section{Limitations and Future Work}
 
Several limitations should be considered. First, the designs lack instruction or data caches, so memory-hierarchy interactions with pipeline depth are outside this study's scope. Second, the branch predictor uses a simple 2-bit saturating counter~\cite{Smith1981} without branch history or target patterns; while identical across all variants and thus not affecting relative comparisons, a more sophisticated predictor would reduce the IPC penalty of deeper pipelines. Third, although RV32 and RV64 variants share identical pipeline control logic, they are not bit-identical, since RV64 requires W-suffix instructions, 6-bit shift-amount decoding, and doubleword memory operations introduce additional control paths. Fourth, the 7-to-8-stage transition was accompanied by twelve RTL-level timing closure optimizations (Section~\ref{sec:timing_closure}), so the frequency improvement at this stage reflects the combined effect of stage insertion and RTL refinement; isolating the contribution of each factor would require evaluating the 8-stage pipeline both with and without the optimizations applied. Fifth, the evaluation does not include instruction-level profiling, so branch-refill, execution-use, and RV64 W-suffix effects are inferred from aggregate trends rather than decomposed quantitatively.
 
Future work will integrate configurable caches, evaluate stronger branch predictors such as a gshare~\cite{McFarling1993} or tournament~\cite{Yeh1991} predictor, and apply instruction-level profiling to decompose individual IPC contributions including EXR insertion, branch-refill latency, execution-use hazards, and W-suffix overhead.
 
\section{Conclusion}
 
This paper presented RV-IM100, a family of ten FPGA-implemented RISC-V microarchitectures derived from a common 5-stage baseline, providing a controlled empirical study that jointly varies datapath width, ISA extension, and pipeline depth within a single processor lineage. Pipeline deepening from 5 to 8~stages raised RV32IM frequency by 192\% (43 to 126\,MHz) and RV64IM by 165\% (38 to 100\,MHz), but the throughput response was non-monotonic: the 6-to-7-stage transition caused absolute throughput regression in RV64 despite higher frequency, while the 7-to-8-stage transition, combined with twelve RTL-level timing closure optimizations, produced the largest throughput improvement in the design space. This non-monotonicity demonstrates that frequency improvement alone is an insufficient predictor of throughput outcome; the IPC cost of a pipeline stage and the frequency headroom available to absorb it must be considered together. Width extension imposed a consistently super-linear resource cost: at 8~stages, RV32 required 59\% fewer LUTs, 51\% fewer FFs, and 80\% fewer DSPs than RV64, while the per-MHz efficiency difference remained modest and benchmark-dependent, with RV64 leading by 2.3\% in DMIPS/MHz and RV32 leading by 4.6\% in CoreMark/MHz. For resource-constrained FPGA targets that do not require 64-bit address space or data width, these results indicate that RV32 offers a substantially more favorable cost-performance trade-off. More broadly, the cross-dimensional data presented here show that width, ISA, and depth extensions interact in ways that cannot be predicted from any single axis in isolation, underscoring the need for joint evaluation when evolving a RISC-V base architecture. All RTL sources and benchmark configurations are released as open-source to support reproducibility and to serve as a reference for the RISC-V microarchitectural design community.\footnote{Available at: \url{https://github.com/T410N/RV-IM100}}

\vfill

\end{document}